\documentclass[twocolumn,english,prl,amssymb,aps,superscriptaddress,showpacs,twocolumn,amsmath,showkeys,floatfix]{revtex4-1}
\usepackage[T1]{fontenc}
\usepackage[latin9]{inputenc}
\usepackage{geometry}
\usepackage[active]{srcltx}
\usepackage{textcomp}
\usepackage{amsmath}
\usepackage{graphicx}
\usepackage{color}
\usepackage{esint}

\makeatletter
\usepackage{babel}

\makeatother

\begin{document}

\title {Mode-Locking of the Hermite-Gaussian Modes of a Nanolaser}
\author{Yifan Sun}
\affiliation{Laboratoire Aim\'e Cotton, Universit\'e Paris-Sud, ENS Paris-Saclay, CNRS, Universit\'e Paris-Saclay, 91405 Orsay Cedex, France}
\author{Sylvain Combrié}
\affiliation {Thales Research and Technology, 91120 Palaiseau, France}
\author{Fabien Bretenaker}
\affiliation{Laboratoire Aim\'e Cotton, Universit\'e Paris-Sud, ENS Paris-Saclay, CNRS, Universit\'e Paris-Saclay, 91405 Orsay Cedex, France}
\affiliation{Light and Matter Physics Group, Raman Research Institute, Bangalore 560080, India}
\author{Alfredo De Rossi}
\affiliation {Thales Research and Technology, 91120 Palaiseau, France}

\begin{abstract} 
Mode-locking is predicted in a nanolaser cavity forming an effective photonic harmonic potential. The cavity is substantially more compact than a Fabry-Perot resonator with comparable pulsing period, which is here controlled by the potential. In the limit of instantaneous gain and absorption saturation, mode-locking corresponds to a stable dissipative soliton\textcolor{black}{, which it very well approximated by} the coherent state of a quantum mechanical harmonic oscillator\textcolor{black}{}. This property is robust against \textcolor{black}{non-instantaneous} material response and non-zero phase-intensity coupling.
\end{abstract}


\maketitle
Mode-locked (ML) diode lasers are compact sources of short pulses with countless applications \cite{Rafailov2007}. The growing relevance of \textcolor{black}{short-distance} and on-chip optical communications for future  computers \cite{Sun2015microprocessor} has stimulated the emergence of novel ultra-compact sources meeting severe energy requirements \cite{Miller2009}. In this respect, a milestone is the achievement of microwatt power consumption with very competitive wall-plug efficiency in nanolaser diodes \cite{Crosnier2017}. A crucial role here is played by the  photonic crystal  (PhC) confining light within a very small volume. These optical sources can be directly modulated \cite{Matsuo2010}. Related to ML is self-pulsing, recently achieved by coupling a PhC cavity with a PhC waveguide Fabry-Perot resonator  \cite{YuMork2017}. 
The aim of this Letter is to introduce a new approach to ML in nanocavities, whereby the eigenfrequencies are engineered to become evenly spaced, as recently demonstrated experimentally \cite{Combrie2017}.
For that purpose, we introduce a model describing the nonlinear laser \textcolor{black}{dynamical behavior} and look for the conditions for robust mode-locking. In particular, we show how the peculiarities of the  mode-locking process in such nanolasers permit to overcome the usual trade-off between laser size and pulsing period. \textcolor{black}{Moreover, we show that the considered ML laser behaves according to the coherent state of a quantum harmonic oscillator, thus bridging the gap between nonlinear nanophotonics and quantum optics.}

\begin{figure}[htbp]
\centering
\includegraphics[width=0.9\columnwidth]{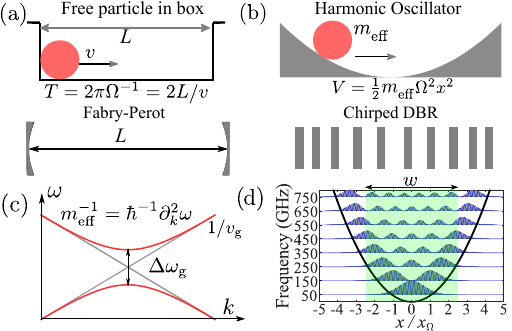}
\caption{
Mechanical analogy of (a) Fabry-Perot resonator and (b) harmonic oscillator implemented in a chirped DBR; (c) Parabolic dispersion in a DBR; (d) Hermite-Gaussian modes in a DBR with parabolic potential. Green area: gain and absorber region.}
\label{Fig01}
\end{figure}

In a Fabry-Perot cavity  (Fig. \ref{Fig01}(a)) and similar resonators (racetracks, whispering gallery modes,..), resonance results from  constructive interference of propagating waves. The round-trip time $T$ is directly related to the resonator length $L$ through $T=2L/v_g$, where $v_g$ is the group velocity. If modes are locked, the cavity round trip time materializes into a pulse propagating back and forth inside the cavity, suggesting a straight mechanical analogy with a free particle bouncing between two barriers. \textcolor{black}{Locking both the longitudinal and transverse modes of a fiber laser has been considered very recently in order to control the spatiotemporal profile of the emitted light \cite{Wise2017}.}

Here, we consider the completely different situation of a photonic resonator made of a metamaterial with effective parabolic dispersion $\omega_{kk}$ and parabolic effective photonic potential $V(x)$ (see Fig.\,\ref{Fig01}(b))\textcolor{black}{, $x$ denoting the spatial degree of freedom}. A laser based on such a resonator cannot be described by  conventional ML laser theory \cite{Haus2000,javaloyes2010}. 
A straightforward implementation of this metamaterial is a quasi-periodic photonic structure. Indeed, Sipe et al.  \cite{Sipe1988,DeSterke1988} have theoretically demonstrated, using  the multiple scales method and Floquet-Bloch theory, that the \textcolor{black}{dynamical behavior} of the slowly varying field envelope $A(x,t)$ of the Bloch waves in a nonlinear medium with periodic dielectric structure is governed by the Nonlinear Schr\"odinger Equation (NLSE). 
The effective parabolic dispersion $\omega_{kk}=\partial^2\omega/\partial k^2$ is the second order dispersion of the normal Bloch mode. In the simplest  periodic structure model (Fig.\,\ref{Fig01}(b)), the coupled mode theory of a Distributed Bragg Reflector (DBR) \cite{Yariv1973} in a waveguide with group velocity $v_g$ relates the dispersion to the width of the photonic band gap $\Delta\omega_g$ through  $\omega_{kk}=2 v_g^2/\Delta\omega_g$.\textcolor{black}{  In general, a PhC allows control of the dispersion \cite{Notomi2001} and in most cases, the band edge is located at a high-symmetry point in the reciprocal space. Consequently, the dispersion is symmetric and third order dispersion is zero. As higher order dispersion could in principle be controlled, this approximation is well justified within the spectral domain of interest (see the detailed discussion in the Supplemental).}

The Gross-Pitaevskii Equation (GPE) is constructed by adding a potential $V(x)$ to the NLSE resulting in a linear confinement \cite{Dobbelaar2015}. 
This can be obtained by building a dielectric guiding nanostructure with one of its parameters, for example the period $a$ of the confining holes, slowly varying with $x$ in a parabolic manner: $a(x)=a_0+\varsigma x^2$. Then, in the limit of small changes of $a$, it can be conjectured that the normal modes and $\omega_{kk}$ are not changed, and the GPE still holds. The only modification is that the local change of $a$ induces a frequency offset of the dispersion $V(x)\propto [a(x)^{-1} - a_0^{-1}]\propto -\varsigma x^2$ \textcolor{black}{ with $\varsigma=-\frac{\Omega^2}{2\omega_{kk}}\frac{a_0}{\omega_c}$ where $\omega_c$ is the edge of the photonic band}. Hence, a chirped periodic dielectric nanostructure results in a harmonic potential $V(x)$ for the field envelope of the normal modes near the band edge. 
A multi-mode high-Q optical resonator with an effective harmonic parabolic potential has been experimentally demonstrated \cite{Combrie2017,suchkov2017}. Such a structure is therefore described by the linear and non-dissipative limit of the GPE equation, written here in a form such that the dispersion $\omega_{kk}$ also appears in the potential term:
\begin{equation}
 \mathrm{i}\frac{\partial A}{\partial t} +\frac{1}{2}\omega_{kk} \frac{\partial^2A}{\partial x^2} -\frac{1}{2}\frac{\Omega^2}{\omega_{kk}}x^2 A =0\ ,
\label{GPE_lin_parab}
\end{equation}
This equation is strictly equivalent to quantum mechanical harmonic oscillator and the envelope is described by a linear superposition of the Hermite-Gaussian eigenmodes  $\Psi_n(x)$ with equally spaced eigenfrequencies $\omega_n=(n +1/2)\Omega+\omega_{c}$\textcolor{black}{}:
\begin{equation}
A(x,t)=\sum_{n=0}^{\infty}C_n(t)e^{-\mathrm{i}\omega_n t}\Psi_n(x)\ .
\label{eq02}
\end{equation}
%
If the modes could be phase locked such as to form the optical equivalent of the coherent state of the quantum harmonic oscillator, then the wavepacket would be described by a Gaussian pulse with velocity and position obeying a sinusoidal evolution without deformation and oscillation period $T=2\pi\Omega^{-1}$ corresponding to the frequency separation $\Omega/2\pi$. The case of the effective harmonic potential is therefore of particular interest in the context of laser.

Hermite-Gauss modes are very different from plane waves as they are spatially inhomogeneous. This has a profound implication in their nonlinear interaction. Moreover, contrary to the Fabry-Perot cavity, \textit{the oscillation period does not depend on the size of the oscillator} but on the effective photon mass $m_{\mathrm{eff}}^{-1}=\hbar^{-1}\partial_k^2 \omega\equiv\hbar^{-1}\omega_{kk}$ of the particle and the stiffness of the potential, which can be expressed as $V(x)=\frac{1}{2}m_{\mathrm{eff}}\Omega^2x^2/\hbar$. The analogy is sketched in Fig. \ref{Fig01}. As the fractional photonic bandgap of PhC cavities is typically $\Delta\omega/\omega\approx20\%$ \cite{Foresi1997} and the group velocity in semiconductor waveguides is about $c_0/4$, \textcolor{black}{the model of a distributed Bragg reflector leads to an order of magnitude estimate of the dispersion  $\omega_{kk}=45\,\mathrm{m}^2\mathrm{s}^{-1}$}. Setting $\Omega/2\pi=100\,\mathrm{GHz}$ leads to a length scale for the Gauss-Hermite modes $x_{\Omega}=\sqrt{\omega_{kk}/\Omega}$, which is equal to $8.4\,\mu\mathrm{m}$. The full width at half maximum of the fundamental mode is then $2\sqrt{ \ln2}\,x_{\Omega}  \simeq 14\,\mu\mathrm{m}$ and the size of a  cavity containing $N$ modes scales as $2\sqrt{N}x_\Omega$ (Fig. \ref{Fig01}(d)). 
 
We investigate now whether such a \textcolor{black}{ comb of modes} can passively mode-lock when this harmonic resonator contains or is hybridized to an active medium providing gain and saturable absorption. The laser nonlinear \textcolor{black}{dynamical behavior} is then described by the Modified GPE (MGPE):
\begin{equation}
 \mathrm{i}\frac{\partial A}{\partial t} +\frac{1}{2}\omega_{kk} \frac{\partial^2A}{\partial x^2} -V(x) A - \mathrm{i} H_1(|A|^2) A=0\ .
\label{eq01}
\end{equation}
$H_1$ holds for the dissipative terms and the nonlinear terms that depend on $|A|^2$:
\begin{equation}
H_1=\frac{1}{2}g(x,t)(1-\mathrm{i}\alpha_{\mathrm{g}}) - \frac{1}{2}a(x,t)(1-\mathrm{i}\alpha_{\mathrm{a}})-\frac{1}{2}\gamma_0\ ,\label{eq03}
\end{equation}
where $g(x,t)$ and $a(x,t)$ are the time and space dependent gain and saturable loss coefficients associated with the Henry factors $\alpha_{\mathrm{g}}$ and $\alpha_{\mathrm{a}}$, respectively, and $\gamma_0$ holds for the intrinsic losses. \textcolor{black}{Here, the gain is assumed spectrally flat, which is realistic for quantum well and quantum dot active materials and a signal bandwidth about 1~THz. }

For simplicity, we first consider the case where  saturation of the gain and losses is instantaneous, leading to:\textcolor{black}{
\begin{equation}
g(x,t)=g_0(x)/\left(1+\frac{{|A(x,t)|^2}}{I_{\mathrm{sat,g}}}\right)\ ,\label{eq:GPE_gain_term_fast}\\
\end{equation}
with a similar expression for $a(x,t)$}. Here $I_{\mathrm{sat,g}}$ and $I_{\mathrm{sat,a}}$ are the saturation intensities for the unsaturated gain and absorption coefficients $g_0$ and $a_0$, respectively. We can then numerically solve eq.\,(\ref{eq01}) for different values of the parameters, and for different values of the widths of the windows into which $g_0$ and $a_0$ are supposed to be homogeneous. A first example is given in Fig.\,\ref{Fig02}, which was obtained when the gain and saturable absorber share the same region of width $w=5x_{\Omega}$ (see Fig.\,\ref{Fig01}\textcolor{black}{(d)}). 
\begin{figure}[]
\centering
\includegraphics[width=0.99\columnwidth]{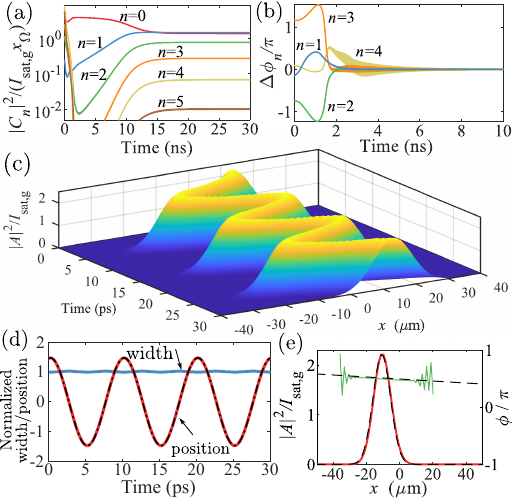}
\caption{Laser behavior for instantaneous gain and absorber saturation with common widths equal to $5x_{\Omega}$. (a,b) Transient evolutions of  (a) normalized intensities and (b) relative phases between the modes after the simulation is started from random mode amplitudes. (c) Evolution of the intracavity intensity in steady-state regime. (d) Positions of the coherent state (black dashed line) and  the soliton (solid red line) and soliton width (solid cyan line) normalized to $x_\Omega$. (e) Amplitude (left axis, solid red line) and phase (right axis, solid green line) of the soliton at a fixed time and corresponding coherent state (dashed line) with amplitude $2.2 I_g$.
%
}
\label{Fig02}
\end{figure}
The parameter values are $\gamma_0=10^{10}\,\mathrm{s}^{-1}$, $r_{\mathrm{g}}=g_0/\gamma_0=5.5$, $r_{\mathrm{a}}=a_0/\gamma_0=9$, and $R_I=I_{\mathrm{sat,g}}/I_{\mathrm{sat,a}}=5$. Such a value of the internal (non saturable) losses $\gamma_0$ is small, but feasible for a semiconductor cavity \cite{Santis2014}. We also suppose here that $\alpha_{\mathrm{g}}=\alpha_{\mathrm{a}}=0$. Eq.\,(\ref{eq01}) is solved starting from random initial conditions, and the behaviors of the different modes are obtained by projecting $A(x,t)$ on the corresponding $\Psi_n(x)$. Figure\,\ref{Fig02}(a) shows the evolution of the normalized intensities of the first six modes, which are the only ones that reach significant steady-state intensities after a few tens of nanoseconds. To determine whether this multimode behavior corresponds to ML operation, we plot the evolutions of the relative phases $\Delta\phi_n=2\phi_n-\phi_{n-1}-\phi_{n+1}$ between the modes for $n=1\ldots 4$ in Fig.\,\ref{Fig02}(b). Here $\phi_n$ is the argument of the mode expansion coefficient  $C_n$ of eq. (\ref{eq02}). One can clearly see that after less than 10~ns all the lasing modes are phase locked. Once steady-state is reached, i.\,e., after about 30~ns, the laser behavior is shown in Fig.\,\ref{Fig02}(c). 

Closer inspection of the spatio-temporal \textcolor{black}{ behavior} reveals a wobbling soliton \cite{grelu2012}, described by sinusoidally varying width and position. These quantities are plotted, normalized to $x_\Omega$, in Fig.\,\ref{Fig02}(d) and coincide almost exactly, except for small residual oscillation ($<1\%$), with the superposition of the linear eigenstates (eq.\,\ref{eq02}), describing the coherent state of the harmonic oscillator in quantum mechanics. This is further apparent in Fig.\,\ref{Fig02}(e) where the amplitude and phase of the field at a fixed time are compared with the coherent state. \textcolor{black}{More detailed analysis is  } given in the Supplemental.

\begin{figure}[]
\centering
\includegraphics[width=0.99\columnwidth]{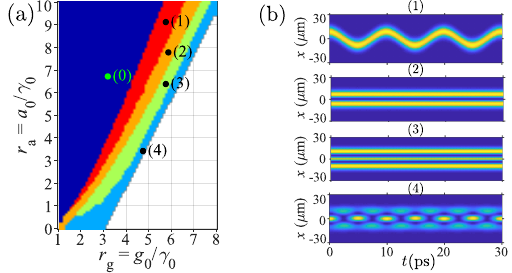}
\caption{(a) Phase diagram: different steady-state regimes versus unsaturated gain and absorption normalized to $\gamma_0$\textcolor{black}{. White hatched region: different unlocked mutimode regimes}. (b) Corresponding false color plots of the laser intensity spatial distribution versus time, for different regimes: (0) below threshold; (1) Soliton-like pulse; (2) Mode $n=1$ alone; (3) Mode $n=2$ alone; (4) Simultaneous oscillation of modes $n=0$ and $n=2$.}
\label{Fig03}
\end{figure}

The nonlinear \textcolor{black}{laser behavior}  undergoes  bifurcations separating different possible behaviors. Colors in Fig.\,\ref{Fig03}(a) represent the regions, in the $\{r_{\mathrm{g}}=g_0/\gamma_0,r_{\mathrm{a}}=a_0/\gamma_0\}$ plane, where different steady-state  behaviors dominate. The soliton-like \cite{grelu2012} pulsed operation of Figs.\,\ref{Fig02}(c-e) does not only require a sufficient amount of gain, but also a sufficient amount of saturable absorption. Our choice for $R_{\mathrm{I}}=I_{\mathrm{sat,g}}/I_{\mathrm{sat,a}}=5$ larger than 1 is also extremely important to obtain this behavior.

\begin{figure}[]
\centering
\includegraphics[width=0.99\columnwidth]{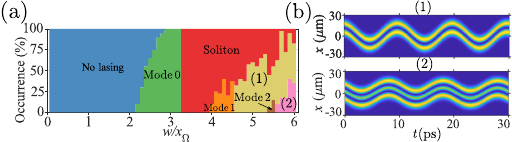}
\caption{(a) Percentage of occurrences of the different regimes, when the simulation is run 40 times with random initial conditions for each value of the gain and saturable absorption window width $w$. Other parameters are $R_{\mathrm{I}}=5$, $g_0=5.5\,\gamma_0$, $a_0=9\,\gamma_0$. (b) False color plot of  laser intensity versus $x$ and $t$ in two examples of regimes labeled (1) and (2) in (a), obtained for  $w=6\,x_\Omega$. They respectively correspond to  oscillation of two or three pulses inside the cavity.}
\label{Fig03N1}
\end{figure}
Although different shapes can be imagined for the gain and absorber, we consider only identical homogeneous gain and saturable absorption windows of width $w$ centered on the potential minimum (green area in Fig.\,\ref{Fig01}(d)). To investigate the influence of $w$, we launch the simulation 40 times, starting from random initial fields, for each value of $w$ ranging from 0 to $6\,x_{\Omega}$. The system exhibits multistability:  it can reach different steady-state regimes for a given set of parameters, depending on the initial values. To gain some statistical insight into this multi-attractor \textcolor{black}{ behavior}, Fig.\,\ref{Fig03N1}(a) displays the occurrences of each regime versus $w$. For $w$ increasing from  $0$ to $3.2\,x_{\Omega}$, the laser is successively below threshold, in single-mode regime, and finally emits the soliton solution of Fig.\,\ref{Fig02}(c). Interestingly, in the range $3.2\,x_{\Omega}\lesssim w\lesssim 3.9\,x_{\Omega}$, soliton emission is the only stable solution. In this example, multistability happens for values of $w$ larger than $3.9\,x_{\Omega}$. For example, for $w=6x_\Omega$, Fig.\,\ref{Fig03N1}(a) shows that the laser \textcolor{black}{dynamical behavior} can fall into three different stable ML regimes, in which one, two, or three pulses oscillate inside the cavity. Figure\,\ref{Fig03N1}(b) gives examples of these two last behaviors. \textcolor{black}{The numerical study reveals that the width $w$ of the gain region is the most important parameter determining the number of locked modes and therefore the spatial amplitude of the pulse oscillation, which is close to $w$. }

\begin{figure}[]
\centering
\includegraphics[width=0.99\columnwidth]{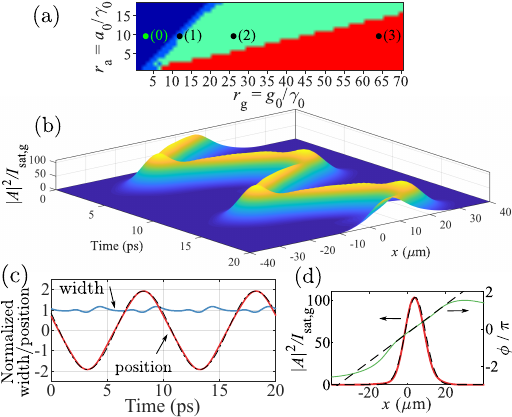}
\caption{Laser behavior for slow gain and absorber saturations. (a) Phase diagram: different steady-state regimes versus unsaturated gain and absorption normalized to $\gamma_0$: (0) below threshold; (1) Single-mode Q-switched operation; (2) Q-switched ML operation; (3) continuous-wave mode-locking. (b,c,d) Stable ML behavior.  Same as Fig.\,\ref{Fig02}(c-e)  corresponding to $r_{\mathrm{g}}=70$ and $r_{\mathrm{a}}=10$ in area number (3).}
\label{Fig04}
\end{figure}
Real semiconductor gain and saturable absorption media have typical response times $\tau_{\mathrm{g}}$ and $\tau_{\mathrm{a}}$ ranging from the picosecond to the nanosecond domain \cite{mecozzi1997}, i. e. not always negligible compared with the photon lifetime in nanocavities or with ps pulse durations like in Fig.\,\ref{Fig02}(c). To investigate the role of such finite lifetimes, we have solved the MGPE eq.\,(\ref{eq01}) with eq.\,(\ref{eq:GPE_gain_term_fast}) replaced by \textcolor{black}{:
\begin{equation}
\frac{\partial g(x,t)}{\partial t} = -\frac{g(x,t)-g_0(x)}{\tau_{\mathrm{g}}} - \frac{|A(x,t)|^2}{\tau_{\mathrm{g}}I_{\mathrm{sat,g}}} g(x,t)\ .
\label{eq:gain_equation}\\
\end{equation}
with a similar equation for $a(x,t)$}. The values of the lifetimes $\tau_{\mathrm{g}}=1\,\mathrm{ns}$ and $\tau_{\mathrm{a}}=10\,\mathrm{ps}$ \cite{Vladimirov2009,Heuck2010}  we choose are those typically mentioned in the literature for InP quantum well lasers  \cite{Jones1995}. Surface recombination has been recently improved in nanostructured lasers owing to advanced passivation techniques such that the carrier lifetime is a few nanoseconds \cite{Crosnier2017} . We also take a ratio of the saturation energies $R_{\mathrm{E}} = I_{\mathrm{sat,g}}\tau_{\mathrm{g}}/I_{\mathrm{sat,a}}\tau_{\mathrm{a}}=25$ from the literature \cite{Heuck2010}. This leads to the results of Fig.\,\ref{Fig04}, computed with $\alpha_{\mathrm{g}}=\alpha_{\mathrm{a}}=0$. The phase diagram of Fig.\,\ref{Fig04}(a) exhibits new regimes, such as passively Q-switched operation, either in unlocked  (point labeled 1) or  Q-switched ML regime (point labeled 2). However, with a proper choice of $a_0$ and $g_0$, one can still obtain cw passively ML operation, as evidenced by the red region of Fig.\,\ref{Fig04}(a). Figures\,\ref{Fig04}(b-d) show one example of such a behavior. The solution is again \textcolor{black}{very close to } a coherent state, although the shape of the pulse gets slightly distorted close to its turning point (see Figs.\,\ref{Fig04}(b,c)). Between these points, Fig.\,\ref{Fig04}(d) shows that the spatial pulse shape is quite well adjusted by a coherent state. Further discussions of discrepancies with respect to a perfect coherent state are provided in the Supplemental. \textcolor{black}{Moreover, a simulation given in Section 5 of the Supplemental indicates that the multi-stability of Fig.\,\ref{Fig03N1} seems to disappear for finite response times, which is positive for practical applications of such nanolasers.}

A non-zero Henry factor, coupling the phase  and intensity variations through  carrier  dynamics, is  known to be a source of instability for passive ML \cite{Vladimirov2005}. However, the laser bifurcation diagram shows that  ML, similar to  Fig.\,\ref{Fig04}, can still be obtained with non zero values of $\alpha_{\mathrm{g}}$ and $\alpha_{\mathrm{a}}$ at the cost of an increase of the pumping (see the Supplemental). 

Experimental implementation of the  harmonic cavity nanolaser  can be envisaged as follows. We consider a photonic crystal made of InP and containing InGaAsP quantum wells to provide enough gain \cite{Matsuo2010,Yu2017} for lasing.  It has also been demonstrated that the laser can be operated well above threshold before any saturation occurs ($I>5I_{th}$ \cite{Crosnier2017}), meaning that the unsaturated gain can exceed many times the non saturable losses. A saturable absorber can be implemented in many ways, for instance like in ref. \cite{Barbay2011}. The harmonic photonic potential is obtained through a suitable design, for instance using a \textcolor{black}{bichromatic} lattice as an alternative to a chirped period \cite{Combrie2017}. As shown in Fig.\,\ref{Fig04}(a), mode-locking could be observed with a gain exceeding saturable losses by a factor close to 5, which is achievable since the Q factor of InP photonic crystals  is about $10^5$ \cite{Crosnier2016}. The large photonic bandgap of these structures results into a large effective photon mass such that the typical cavity size for  100 GHz  period would be less than $100\,\mu\mathrm{m}$. Thus, the essential requirements for building an extremely compact ML integrated nanolaser are met by the current state-of-the-art of  nanolaser technology. \textcolor{black}{Balanced extraction of power from all the locked modes could be achieved with a geometry like in  \cite{Crosnier2017}}.

In conclusion, a novel concept for ML in ultracompact semiconductor lasers has been proposed, based on a harmonic potential to confine light. This maps the optical cavity into a quantum mechanical harmonic oscillator, with evenly spaced eigenfrequencies, an essential requirement for ML. 
The nonlinear \textcolor{black}{ behavior} is described by the Gross-Pitaevskii equation with a parabolic potential and nonlinear terms describing  gain and absorption. ML occurs with Hermite-Gaussian modes, which are very different from  waves of usual resonators, as they are stationary modes with a strongly inhomogeneous spatial distribution of  energy. Provided that saturable gain and absorption overlap with all the modes, ML occurs over a broad area in the phase space, corresponding to the emergence of dissipative soliton and multi-soliton solutions. In the limit of instantaneous absorption and gain saturation, the dissipative soliton is well described by the coherent state of a quantum mechanical oscillator, namely a Gaussian envelope oscillating without deformation. ML period is controlled by the design of the photonic potential, and not by the  cavity length. For a fixed ML period, here 10~ps, the linear size of our cavity is $80\,\mu\mathrm{m}$, about five times more compact than for a Fabry-Perot laser  made with the same material. Finally, slow absorption/gain \textcolor{black}{response} still allows ML and most features of the coherent state are retained. Thus, the concept of ML based on Hermite-Gaussian modes in photonic nanostructures could solve the \textcolor{black}{long-standing} problem of miniature periodic pulsed sources. \textcolor{black}{Moreover, the laser behaving like a quantum harmonic oscillator makes a link between nonlinear nanophotonics and quantum optics, opening the way to interesting ramifications regarding quantum photon statistics in such ML nanolasers. Furthermore, analyzing the full bifurcation diagram \cite{Gurevich2019} is interesting  from the viewpoint of nonlinear dynamical systems theory.}

\begin{acknowledgments}
Work supported by the  Direction G\'en\'erale de l'Armement (LASAGNE, ANR-16-ASTR-0010-03), the ``Investissements d'Avenir'' program (CONDOR, ANR-10-LABX-0035), European Union's Horizon 2020  program  (Fun-COMP, grant agreement  780848) and performed in the framework of the joint research lab between TRT and LAC. 
\end{acknowledgments}

\newpage

{\huge Supplementary information}
%
%

\section{Dispersion and potential in a PhC cavity}
\begin{figure}[h]
\centering
\includegraphics[width=0.5\columnwidth]{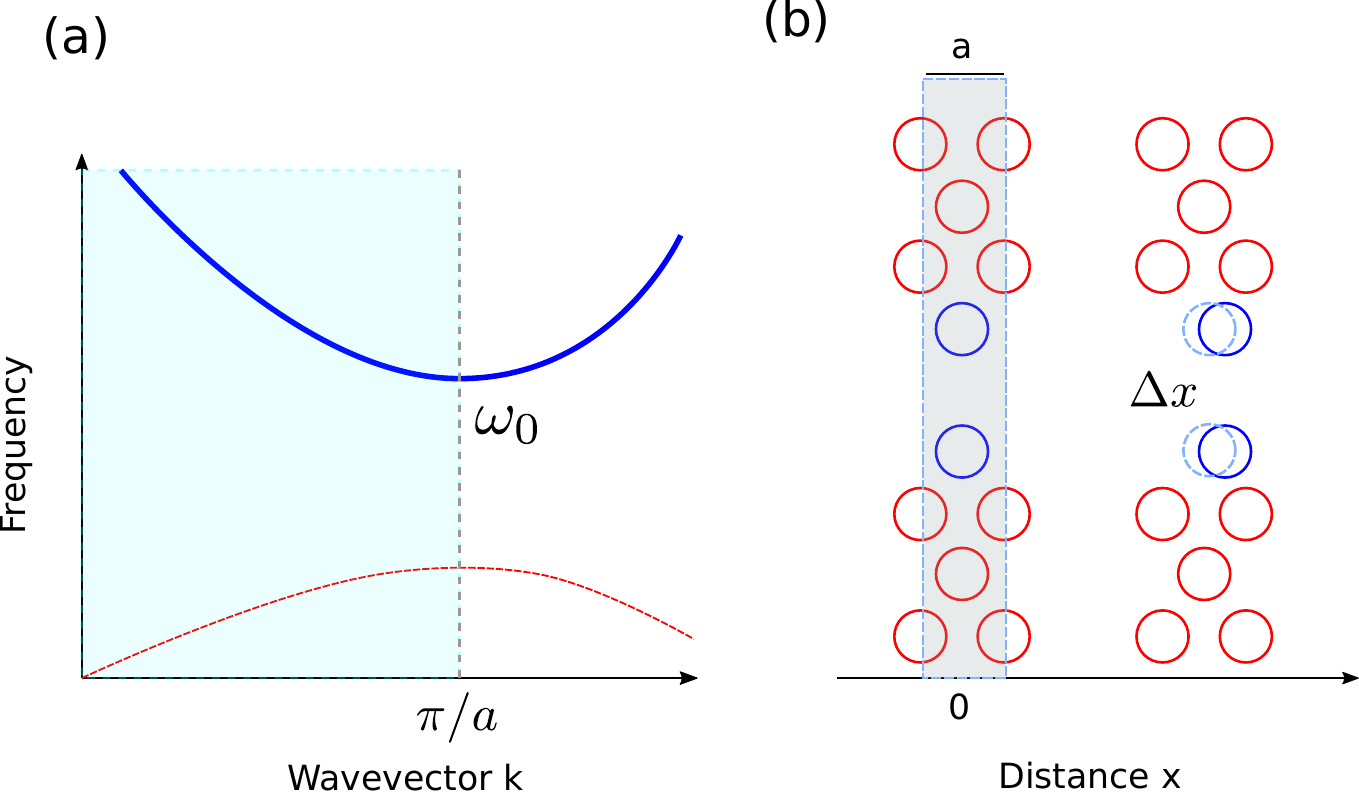}
 \caption{\label{fig:dispersion_concept}(a) Typical dispersion of a photonic crystal waveguide; the localized guide mode corresponds to the blue band; (b) geometry of the bichromatic lattice, the periodic cell is in grey, the lattice with period $a^\prime$ is represented in blue.}
 \end{figure}

\begin{figure}[]
\centering
\includegraphics[width=0.6\columnwidth]{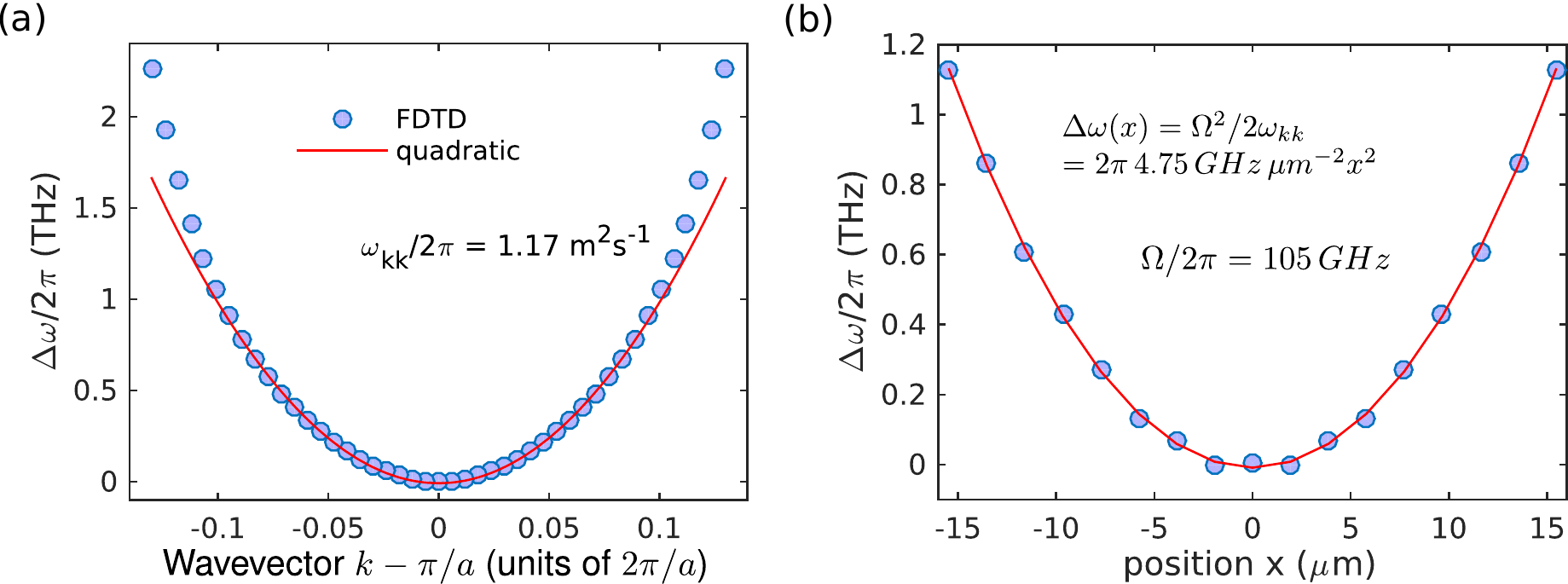}
\caption{\label{fig:dispersion}(a) Calculated dispersion for a Bloch-periodic PhC; frequency offset relative to the band edge $\Delta\omega=\omega-\omega_0$ vs. normalized wavevector near the edge of the first Brillouin zone ($\pi/a$), quadratic fit (solid line); (b) dependence of the band edge $\Delta\omega = \omega_0 -\min{(\omega_0)}$ on the position in the cavity and parabolic fit (solid line).}
\end{figure}

The purpose of this Section is to describe the ``effective photonic potential'' in terms of the spatial dependence of the dispersion.
Let us consider the case of the bichromatic cavity as described in \cite{Combrie2017}. The structure, shown in Fig.\,\ref{fig:dispersion_concept}(b), consists of a waveguide oriented along $x$ with broken periodicity as the period $a^\prime$ of the inner row of holes is different from the period of the lattice $a$. As a consequence, at each lattice period, the inner holes are shifted by $\Delta x=(1-a^\prime/a) x$, where $x$ is the position of the section of the crsytal. Let us consider the photonic band of a strictly periodic structure when $\Delta x=0$ and calculate the dispersion of the corresponding Bloch mode. The band considered is that which is waveguided (see Fig. \ref{fig:dispersion_concept}(a)).\\
The dispersion is calculated using the Finite Difference in Time Domain over a cell with period $a$ (see Fig. \ref{fig:dispersion_concept}(b)), using Floquet-Bloch boundary conditions $u(x+a)=u(x)\exp(\mathrm{i} k a)$. The parameters are: slab thickness = 180 nm, period $a$= 485 nm, radius of the holes $0.27\;a$, refractive index 3.17. 
The result is shown in Fig . \ref{fig:dispersion}(a). The quadratic fit $\omega = \omega_0 + \omega_{kk}k^2/2$ is very good within a band of 1 THz from the bottom of the band, with $\omega_{kk}/2\pi=1.17\;\mathrm{m}^2\mathrm{s}^{-1}$. Higher order dispersion is noticeable only above 1 THz. 
Thus, the propagation of Bloch waves is well described by the Schr\"{o}dinger equation:
\begin{equation}
 \mathrm{i}\frac{\partial A}{\partial t} +\frac{1}{2}\omega_{kk} \frac{\partial^2A}{\partial x^2} -\frac{1}{2}\frac{\Omega^2}{\omega_{kk}}x^2 A =0\ .
\label{GPE_lin_parab}
\end{equation}
As the relative position of the holes changes by $\Delta x$, the lower edge of the dispersion is changed. When $a^\prime/a=0.995$, this corresponds to a potential $V(x)$ which is approximately parabolic, namely $V(x)=\Omega^2/2\omega_{kk} x^2$. From the fit $\Omega^2/4\pi\omega_{kk}=4.75\,\mathrm{GHz}.\,\mu\mathrm{m}^2$, therefore the spectral separation of the eigenmodes is $\Omega/2\pi=105$ GHz. We note that $\Omega$ scales as $1-a^\prime/a$, therefore, when $a^\prime/a=0.98$ as in \cite{Combrie2017}, then $\Omega/2\pi=420$ GHz, which is very close to the experimental results.
Thus, the Schr\"{o}dinger equation describes reasonably well the confinement in a certain class of photonic crystal cavities, as long as eigenmodes close enough to the photonic band edge are considered.

\section{Comparison between stable soliton state and coherent state}
\begin{figure}[h]
	\centering
	\includegraphics[width=0.4\columnwidth]{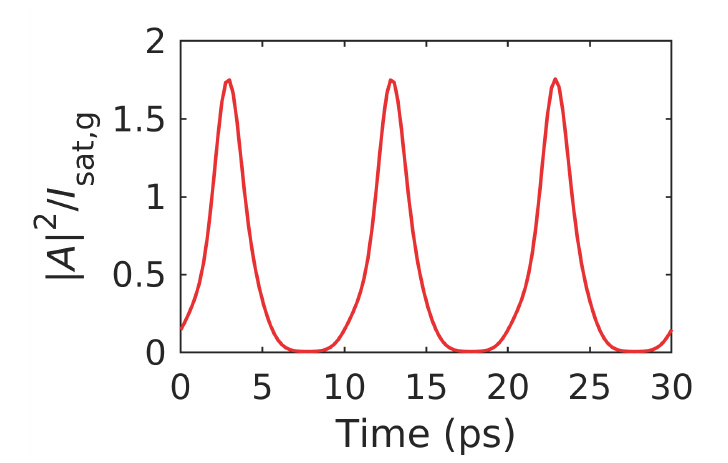}
	\caption{Time evolution of the intensity of the pulse travelling in the $+x$ direction at $x = 0$ in the case of the solution corresponding to Fig. 2(c-e) in the Letter.}
	\label{FigPulseTemporel1}
\end{figure}

A more detailed comparison between the soliton solution of the MLSE of Fig. 2 in the Letter with the coherent state  quantum mechanical harmonic oscillator is carried out here. This completes the comparison already performed in Figs. 2(d) and 2(e) of the Letter. First, we reproduce in Fig.\;\ref{FigPulseTemporel1} the time evolution of the pulse propagating at $x=0$ in the $+x$ direction by keeping only the positive values of $k$ in the spatial Fourier domain, in the case the solution of Figs. 2(c-e) of the Letter. The pulse has a duration of 2.3~ps.
The field in a coherent state can be expressed as the superposition of Hermite-Gaussian modes:
\begin{equation}
A\mathrm{_{coh}}(x,t)=\sum_{n=0}^{\infty}C_ne^{-\mathrm{i}(n+\frac{1}{2})\Omega t}\Psi_n(x)\ ,
\label{eq02}
\end{equation}
where $\Psi_n(x)$ is the Hermite-Gaussian modes of order $n$ and $\Omega$ the free spectral range. For a coherent state, the mode intensities $|C_n|^2$ follow a Poisson distribution with parameter $\lambda$:
\begin{equation}
|C_n|^2 = \frac{\lambda^n}{n!}e^{-\lambda}.
\end{equation}
The expression of the normalized field in such a coherent state can be re-written as
\begin{equation}
A\mathrm{_{coh}}(x,t)= \pi^{-1/4}x_\Omega^{-1/2}e^{-\frac{1}{2}\left(\frac{x-\sqrt{2\lambda}x_\Omega\cos(\Omega t)}{x_\Omega}\right)^2+\mathrm{i}\phi(x,t)}\ ,
\label{field_int}
\end{equation}
with
\begin{equation}
\phi(x,t)= -\frac{\sqrt{2\lambda}\sin(\Omega t)}{x_\Omega}x-\frac{1}{2}\Omega t+\frac{1}{2}\lambda\sin(2\Omega t)\ .
\label{field_phase}
\end{equation}
The quantity $x_\Omega$ is the scaling factor defined in the Letter.
\begin{figure}[h]
	\centering
	\includegraphics[width=0.6\columnwidth]{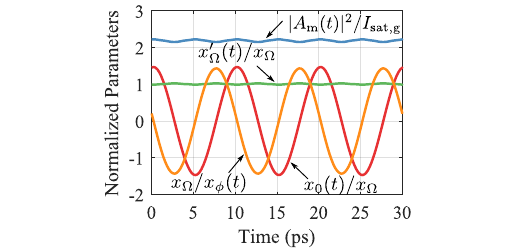}
	\caption{Time dependence of the normalized parameters $x'_\Omega(t)/x_\Omega$, $x_\Omega/x_\phi(t)$, $x_0(t)/x_\Omega$, $A_m(t)/I_\mathrm{sat,g}$ of eqs. \eqref{field_int}  and \eqref{field_phase}
such that these equations match the stable soliton state in Fig. 2(d) in the Letter. $x_\Omega = \sqrt{\omega_{kk}/\Omega}=8.4346\mu m$
	}
	\label{Fig02}
\end{figure}

To compare the coherent state with the final stable soliton state of Fig. 2 of the Letter, the numerical solution of the MGPE is modeled with 
the function:
\begin{equation}
A\mathrm{_{fit}}(x,t)= A_{\mathrm{m}}(t)e^{-\frac{1}{2}\left(\frac{x-x_\mathrm{o}(t)}{x'_\Omega(t)}\right)^2+\mathrm{i}\phi\mathrm{_{fit}}(x,t)}\,, \label{field_int}\ 
\end{equation}
with
\begin{equation}
\phi\mathrm{_{fit}}(x,t)= \frac{x}{x_\phi(t)}+\phi_{\mathrm{extra}}(t)\ .
\label{field_phase}
\end{equation}
Through best fit at each time step, the parameters $A_{\mathrm{m}}(t)$, $x_\mathrm{o}(t)$, $x'_\Omega(t)$,  and $x_\phi(t)$ are extracted and plotted as a function of time in  Fig.\;\ref{Fig02}, which corresponds to Fig. 2(d) in the Letter. Within the considered 30\,ps time window, we obtain a value of $x'_\Omega(t)/x_\Omega= 1.009\pm0.01 \mu\mathrm{m}$ and $|A_{\mathrm{m}}(t)|^2/I_\mathrm{g}=2.189\pm  0.024$. Within the uncertainty corresponding to the small fluctuations around the average, this shows that the soliton coincides with a coherent state, as, in addition to that, all the parameters of the model function of eq.\, \eqref{field_int} follow the prescribed time dependence.

\begin{figure}[h]
	\centering
	\includegraphics[width=0.4\columnwidth]{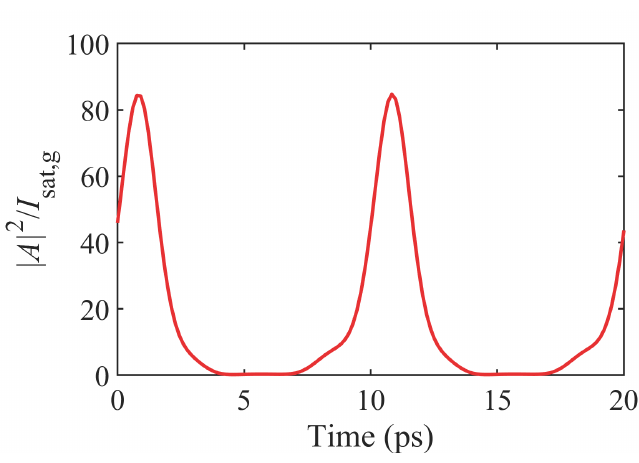}
	\caption{Time evolution of the intensity of the pulse travelling in the $+x$ direction at $x = 0$ in the case of the solution corresponding to Fig. 5(b-d)) in the Letter.}
	\label{FigPulseTemporel2}
\end{figure}
\begin{figure}[h]
	\centering
	\includegraphics[width=0.6\columnwidth]{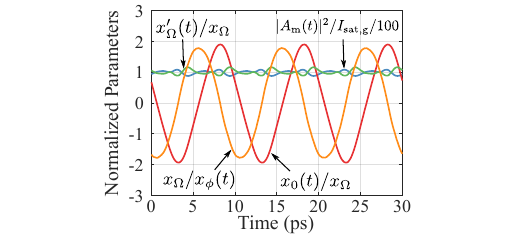}
	\caption{Time dependence of the normalized parameters $x'_\Omega(t)/x_\Omega$, $x_\Omega/x_\phi(t)$, $x_0(t)/x_\Omega$, $A_m(t)/I_\mathrm{sat,g}$ of eqs. \eqref{field_int}  and \eqref{field_phase}
such that these equations match the stable soliton state in Fig. 5(b-d) in the Letter. $x_\Omega = \sqrt{\omega_{kk}/\Omega}=8.4346\mu m$
	}
	\label{Fig02bis}
\end{figure}

Similarly, Figs.\;\ref{FigPulseTemporel2} and \ref{Fig02bis} reproduce the pulse propagating at $x=0$ and the parameters extracted from the results plotted in Fig. 5(b-d) of the Letter, corresponding to simulations performed with non instantaneous gain and absorption saturation. The pulse duration is now 1.8~ps. Within the considered 30\,ps time window, they correspond to $x'_\Omega(t)= 8.4603\pm 0.65385 \mu\mathrm{m}$ and $|A_{\mathrm{m}}(t)|^2/I_\mathrm{g}=99.2919\pm  6.0582$. Although the parameters fluctuate a bit more than in the case of instantaneous saturation, the solution is still very close to the coherent state of a quantum harmonic oscillator.

\section{Numerical solving of the Modified Gross-Pitaevskii Equation}
The Modified Gross-Pitaevskii Equation (MGPE) is solved using the Fourier split-step method. The space grid contains $N=2^m$ nodes and defined as $x_n=n\Delta x$, with $n\in[-N/2, N/2-1]$. The points in the reciprocal space are $k_n=2\pi N n/\Delta x$. The MGPE is recast in the form:
\begin{equation}
\partial_t A = \mathbf{D} A + \mathbf{N}A\ .
\end{equation}
$\mathbf{D} A$ is the dispersion term in the equation, and $\mathbf{N}A$ holds for the the other terms. The dispersion operator is calculated in the reciprocal space by using the spatial fast Fourier transform (FFT) of the field: 
\begin{equation}
\mathbf{D} A = \mathrm{FFT}^{-1}[\frac{\mathrm{i}}{2}\omega_{kk}k^2 \mathrm{FFT}(A) ]\ .
\end{equation}
The other term are calculated in the real space and finally the time evolution of the field is calculated by a generic numeric ordinary differential equation solver.
\section{Influence of a non-zero Henry factor}
\begin{figure}[]
\centering
\includegraphics[width=0.6\columnwidth]{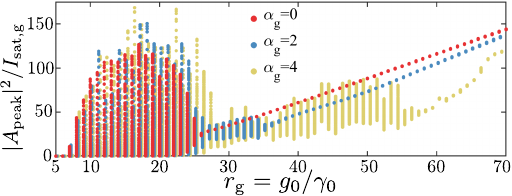}
\caption{Bifurcation diagram for different values of $\alpha_{\mathrm{g}}$, with $\alpha_{\mathrm{a}}=\alpha_{\mathrm{g}}$. The values of the peak intensities at the center of the cavity are shown as a function of $r_{\mathrm{g}}=g_0/\gamma_0$. For each point, the simulation is run till the system reaches steady-state.  The values of the other parameters are $r_{\mathrm{a}} = a_0/\gamma_0=5$ and the gain and absoption window width is equal to $5x_{\Omega}$, like in Figs. 2, 3, and 4 of the Letter.}
\label{Fig05}
\end{figure}
As stated in the Letter, a non-zero Henry factor in the gain section or in the saturable absorber can be a source of instability for passive mode-locking. To investigate this possibility, Fig. \ref{Fig05} shows a bifurcation diagram obtained by varying the laser unsaturated gain $g_0$, all other parameters being kept fixed. For each value of $g_0$, the simulation of the MGPE (eq.\,3 of the Letter) is run till the laser reaches a steady-state behavior. Then, the subsequent evolution of the laser is simulated for a further duration of 5~ns. After that, all the intensity peaks that occur at $x=0$ during this 5-ns-long sample are detected and their peak intensities are plotted on the vertical scale of the diagram. Figure\,\ref{Fig05} reproduces three bifurcation diagrams, corresponding to  $\alpha_{\mathrm{g}}=\alpha_{\mathrm{a}}=0,2,4$. In each case, increasing $g_0$ leads to a transition from a regime in which the emission is not $2\pi/\Omega$-periodic (several points on the same vertical line in the bifurcation diagram), to a regime where it becomes periodic (only one point per vertical line). The first regime corresponds to Q-switched ML operation and the second one to a continuous ML soliton-like emission regime very similar to those of Figs.\,\ref{Fig02}(c) and \ref{Fig02}(d). Figure\,\ref{Fig05} also clearly shows that a large Henry factor makes mode-locking more unstable. For example, the value of $r_{\mathrm{g}}=g_0/\gamma_0$ over which stable mode-locking is obtained increases from about 26 for $\alpha_{\mathrm{g}}=\alpha_{\mathrm{a}}=0$  to 36 for $\alpha_{\mathrm{g}}=\alpha_{\mathrm{a}}=2$ to 58 for $\alpha_{\mathrm{g}}=\alpha_{\mathrm{a}}=4$. However, in all cases, stable soliton emission can be obtained provided one is able to operate the laser far enough from threshold.

We have also verified numerically that this conclusion still holds when $\alpha_{\mathrm{g}}\neq\alpha_{\mathrm{a}}$.
\section{Influence of the non-instantaneous response of the gain and absorber media on the laser multi-stability}
\begin{figure}[]
\centering
\includegraphics[width=0.6\columnwidth]{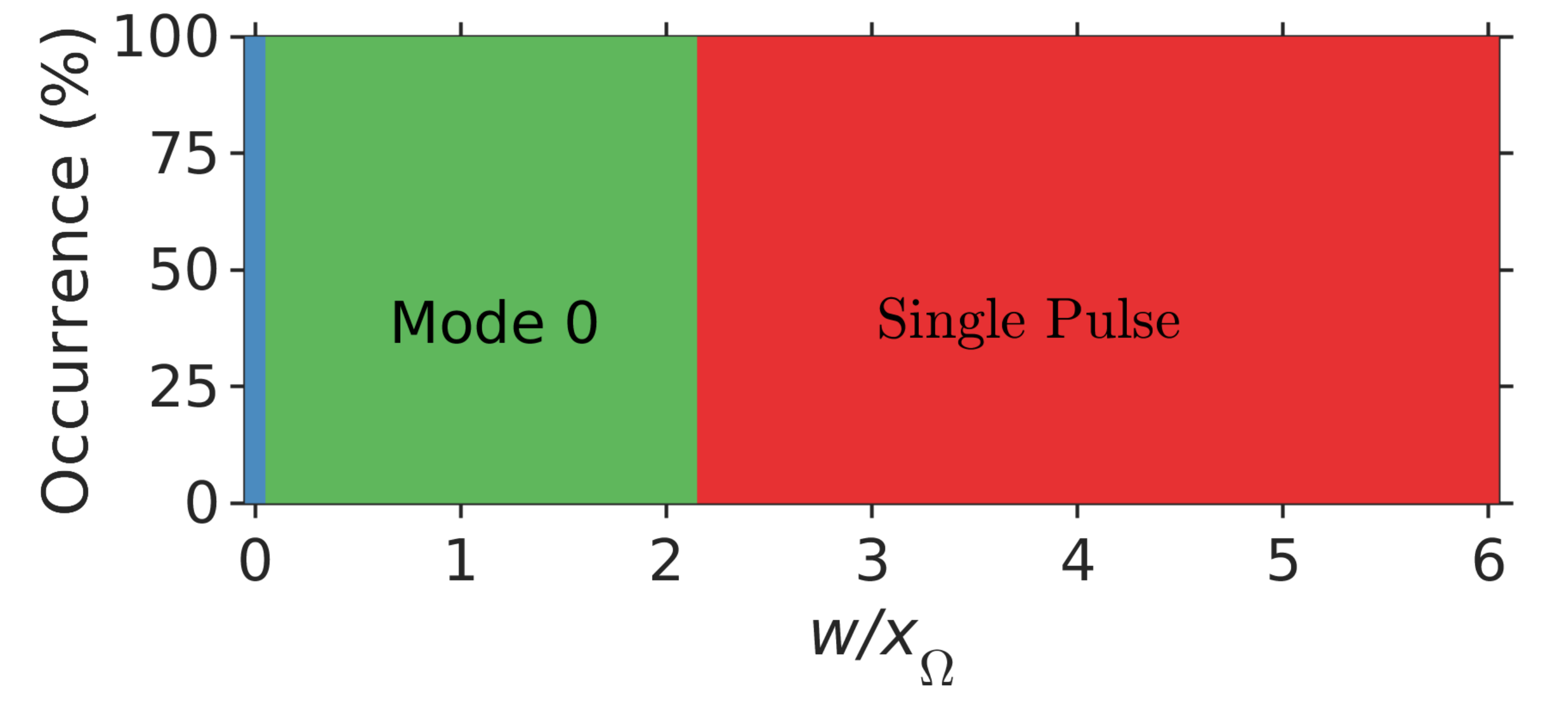}
\caption{Same figure as Fig. 4(a) of the Letter for slow gain and absorber saturations. The parameters have the same values as in Fig. 5(b-d) of the Letter. }
\label{FigSlowBifurc}
\end{figure}
Figure 4(a) of the Letter shows that the laser can exhibit multi-stability, meaning that the steady-state regime reached by the laser can be sensitive to initial conditions. Although this is interesting from the point of view of the study of dynamical systems, this can be an issue for applications. The question then arises to know whether this multi-stability gets more pronounced when one takes into account the finite response time of the gain and absorber. An element of response is given in Fig.\,\ref{FigSlowBifurc}, which was obtained with the same values of the parameters as in Fig. 5 of the Letter, except of course for the width $w$ of the active region. One can see that the multis-stability completely disappears: the laser becomes fully deterministic, which is very encouraging for practical applications.

\end{document}